\documentclass[epj]{svjour}
\usepackage{graphics}
\usepackage{amssymb}
\begin{document}

\title{Double-exchange model study of multiferroic $R$MnO$_3$ perovskites}

\author{Shuai Dong\inst{1,2,3} \and Rong Yu\inst{1,2} \and Seiji Yunoki\inst{4,5} \and J.-M. Liu\inst{3,6} \and Elbio Dagotto\inst{1,2}}

\institute{Department of Physics and Astronomy, University of Tennessee, Knoxville, TN 37996, USA \and
Materials Science and Technology Division, Oak Ridge National Laboratory, Oak Ridge, TN 32831, USA \and
Nanjing National Laboratory of Microstructures, Nanjing University, Nanjing 210093, China \and
Computational Condensed Matter Physics Laboratory, RIKEN, Wako, Saitama 351-0198, Japan \and
CREST, Japan Science and Technology Agency (JST), Kawaguchi, Saitama, 332-0012, Japan \and
International Center for Materials Physics, Chinese Academy of Sciences, Shenyang 110016, China}

\date{Received: Jan. 25, 2009}

\abstract{In this proceeding, recent theoretical investigations by the authors on the multiferroic $R$MnO$_3$ perovskites are briefly reviewed at first. Using the double-exchange model, the realistic spiral spin order in undoped manganites such as TbMnO$_3$ and DyMnO$_3$ is well reproduced by incorporating a weak next-nearest neighbor superexchange ($\sim10\%$ of nearest neighbor superexchange) and moderate Jahn-Teller distortion. The phase transitions from the A-type antiferromagnet (as in LaMnO$_3$), to the spiral phase (as in TbMnO$_3$), and finally to the E-type antiferromagnet (as in HoMnO$_3$), with decreasing size of the $R$ ions, were also explained. Moreover, new results of phase diagram of the three-dimensional lattice are also included. The ferromagnetic tendency recently discovered in the LaMnO$_3$ and TbMnO$_3$ thin films is explained by considering the substrate stress. Finally, the relationship between our double-exchange model and a previously used $J_1$-$J_2$-$J_3$ model is further discussed from the perspective of spin wave excitations.
}
\maketitle

\section{Introduction}
Recently, the multiferroic materials, in which the ferroelectric (FE) and magnetic orders coexist and are intimately coupled, have attracted much attention due to their technological relevance and fundamental science challenges \cite{Fiebig:Jpd,Eerenstein:Nat,Cheong:Nm,Ramesh:Nm}. Among the single phase multiferroic materials, the undoped manganites with small size $R$ (rare-earth) cations (like Tb, Dy, Ho and so on) is one of the most fascinating families. Not only the perovskites or hexagonal $R$MnO$_3$, BiMnO$_3$, and YMnO$_3$ \cite{Kimura:Nat,Goto:Prl,Kenzelmann:Prl,Kimura:Prb05,Arima:Prl,Lorenz:Prb}, but also the $R$Mn$_2$O$_5$ series \cite{Hur:Nat,Chapon:Prl,Cruz:Prb} show multiferroicity.

In this work, we only consider the undoped perovskite-type $R$MnO$_3$. With decreasing $R$ size, the ground state of these $R$MnO$_3$ compounds changes from the A-type antiferromagnet (A-AFM) (like LaMnO$_3$ and NdMnO$_3$), to the spiral spin state (like TbMnO$_3$ and DyMnO$_3$), and then finally to the E-AFM (like HoMnO$_3$) \cite{Goto:Prl,Zhou:Prl}, which is referred below as the ``A-S-E transition''. On one hand, the spiral spin order (SSO) can break the space inversion symmetry and thus induce the observed ferroelectric (FE) polarization \cite{Kimura:Nat,Goto:Prl,Kenzelmann:Prl,Kimura:Prb05,Arima:Prl,Mostovoy:Prl}, although its microscopic mechanism remains under debate \cite{Katsura:Prl,Sergienko:Prb,Xiang:Prl,Malashevich:Prl}. This SSO driven improper ferroelectricity is also observed in other materials \cite{Kimura:Armr}. On the other hand, the E-AFM spin order can also induce the FE polarization because its zigzag chains break the space inversion symmetry as well \cite{Lorenz:Prb,Sergienko:Prl,Picozzi:Prl,Yamauchi:Prb}.

In contrast to the E-AFM phase which can be easily obtained using the two-orbital double-exchange (DE) model \cite{Hotta:Prl}, the origin of the SSO remains a puzzle. A direct but phenomenological route to generate a SSO phase is via the magnetic frustration between NN ferromagnetic (FM) and next-nearest-neighbor (NNN) antiferromagnetic (AFM) interactions \cite{Cheong:Nm}, e.g. via a $J_{1}$-$J_{2}$-$J_{3}$ model with classical spins, where $J_{1}$ is the NN superexchange (SE) while $J_{2}$ ($J_{3}$) is the NNN SE along the $b$ ($a$) directions \cite{Kimura:Prb}. Another route to obtain the SSO phase is to incorporate the Dzyaloshinskii-Moriya (DM) interaction ($\propto \textbf{S}_i\times \textbf{S}_j$) into the DE framework \cite{Sergienko:Prb,Li:Prb}. However, these models are not sufficient to describe the several phases of $R$MnO$_3$. Instead, our recent work has proposed an alternative model to understand the SSO and A-S-E transition in $R$MnO$_3$ \cite{Dong:Prb08.2}, which will be the focus of this manuscript.

The rest of paper is organized as follow: In Sec.~2, we introduce the DE model and calculation methods. In Sec.~3, our recent study of the SSO and phase diagram in the two-dimensional (2D) lattice is briefly reviewed \cite{Dong:Prb08.2}. In Sec.~4, the study of the phase diagram is extended to the three-dimensional (3D) lattice. In Sec.~5, discussions regarding the spin-wave spectrum of SSO are presented. The main conclusions are summarized in Sec.~6.

\begin{figure}
\centering \resizebox{0.5\textwidth}{!}{\includegraphics{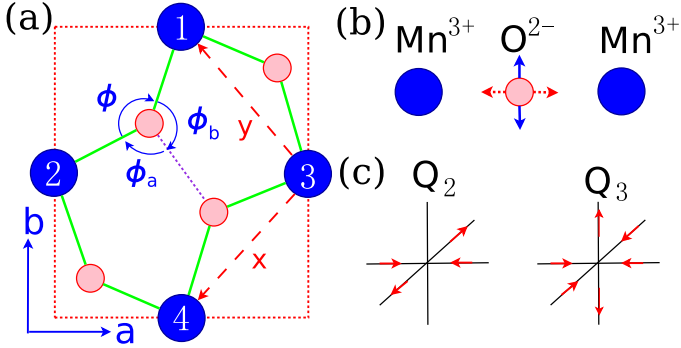}} \caption{(Color online) (a) Sketch of the crystal structure ($a-b$ plane) of $R$MnO$_{3}$. Two types of coordinate axes ($a-b$ and $x-y$) are shown. (b) Illustration  of the two kinds of distortions discussed in the text: GdFeO$_{3}$-type (oxygen moves perpendicular to the Mn-O-Mn bond) and Jahn-Teller type (oxygen moves along Mn-O-Mn bond). (c) Two Jahn-Teller distortion modes: $Q_{2}$ and $Q_{3}$. Reproduced from \cite{Dong:Prb08.2}. Copyright 2008, the American Physical Society.}
\end{figure}

\section{Model and Methods}
Theoretically, the physics of manganites can be qualitatively understood within the framework of the DE model, including the SE between the $t_{\rm 2g}$ spins, and the Jahn-Teller (JT) interaction \cite{Dagotto:Prp}. To obtain the spiral spin order, a weak NNN SE also has to be considered. The whole Hamiltonian reads as:
\begin{eqnarray}
\nonumber
H&=&-\sum_{<ij>}^{\alpha\beta} t_{\textbf{r}}^{\alpha\beta}\Omega_{ij} c_{i\alpha}^{\dagger} c_{j\beta} + J_{\rm AF} \sum_{<ij>}\textbf{S}_{i} \cdot \textbf{S}_{j}\\
\nonumber
&&+\sum_{[ik]}J_{2\gamma}\textbf{S}_{i}\cdot\textbf{S}_{k}+\lambda\sum_{i}[Q_{2,i}\tau_{x,i}+Q_{3,i}\tau_{z,i}]\\
&&+\frac{1}{2}\sum_{i}(Q_{2,i}^2+Q_{3,i}^2),
\end{eqnarray}
where the first term is the standard DE of the $e_{\rm g}$ electrons. The DE hopping amplitudes $t_{\textbf{r}}^{\alpha\beta}$ are orbital- and direction-dependent. In particular, $t_{x}^{1,1}=t_{y}^{1,1}=3t_{x}^{2,2}=3t_{y}^{2,2}=\frac{3}{4}t_{0}$, $t_{y}^{1,2}=t_{y}^{2,1}=-t_{x}^{1,2}=-t_{x}^{2,1}=\frac{\sqrt{3}}{4}t_{0}$, $t_{z}^{1,1}=t_{z}^{1,2}=t_{z}^{2,1}=0$, and $t_{z}^{2,2}=t_{0}$ where the superscript $1$ ($2$) denotes the $e_{\rm g}$ orbital $d_{x^2-y^2}$ ($d_{3z^2-r^2}$), and $t_{0}$ ($\sim0.2-0.3$ eV) is taken as the energy unit \cite{Dagotto:Prp}. The infinite Hund coupling used here generates a Berry phase $\Omega_{ij}=\cos(\theta_i/2)\cos(\theta_j/2)+\sin(\theta_i/2)\sin(\theta_j/2)\exp[-i(\varphi_i-\varphi_j)]$, where $\theta$ and $\varphi$ are the angles defining the $t_{\rm 2g}$ spins $\textbf{S}$ in spherical coordinates. The second term in the Hamiltonian is the usual AFM SE coupling between NN $t_{\rm 2g}$ spins. A realistic $J_{\rm AF}$ value is about $\sim10\%t_0$ according to previous studies \cite{Dagotto:Prp}. The third term is the NNN SE between $3d$ spins, where $\gamma$ is the direction index. $J_2$ is anisotropic due to the lattice distortion, e.g.  the NNN SE $J_{\rm 2b}$ coupling between Mn(1) and Mn(4) can be $\sim1.8-2.2$ times the value of the $J_{\rm 2a}$ coupling between Mn(2) and Mn(3) since $b>a$ \cite{Dong:Prb08.2}, as sketched in Fig.~1(a). The fourth term is the electronic-phonon coupling of JT distortion, where $\lambda$ is the spin-phonon coupling coefficient and $\tau$ is the orbital pseudospin operator, given by $\tau_{x}=c_a^{\dagger}c_b+c_b^{\dagger}c_a$ and $\tau_z=c_a^{\dagger}c_a-c_b^{\dagger}c_b$ \cite{Dagotto:Prp}. For all $R$MnO$_3$ materials at low temperatures, $|Q_2|$ and $Q_3$ are uniform with $|Q_2|\approx-\sqrt{3}Q_3$. The sign of $Q_2$ is staggered, which gives rise to the well-known staggered $d_{\rm 3x^2-r^2}$ and $d_{\rm 3y^2-r^2}$ orbital ordering \cite{Zhou:Prb}. The last term is the elastic energy of JT phonons.

The numerical method used in this manuscript, and in the cited related references by our group, is the variational method employed at zero temperature (zero-T). The total energies (per site) of several candidate phases are compared to determine which is the most likely ground state. The 2D candidate phases include: A-AFM, C-AFM, CE, C$_{1/4}$E$_{3/4}$, C$_{1/3}$E$_{2/3}$, Dimer, E-AFM, G-AFM, SSO state \cite{Hotta:Prl,Dong:Prb08.2,Dagotto:Prp}. In this list almost all the 2D typical spin order patterns discussed in manganites have been included. For the SSO, spirals with wave vectors $q$ from $0$ to $1/4$ are taken into account \cite{note}. For the 3D calculation, the FM phase and canting spin state are added to the candidate list. In principle, the $t_{\rm 2g}$ spins $\textbf{S}$ in Eq.(1) can be Heisenberg-like. However, in real manganites, all spin patterns of the known magnetically ordered phases can be described using the X-Y model, namely considering the existence of an easy magnetic plane, such as the $b$-$c$ plane for the SSO in TbMnO$_3$ and DyMnO$_3$. Therefore, in practice, all the candidate phases considered here have X-Y model like spin patterns. Once a spin pattern and a JT distortion are selected, the total energy (per site) can be calculated in the \emph{infinite} size lattice limit (thus, there are no finite-size effects here). This energy 
includes: (1) the DE kinetic energy $E_{\rm K}$ (including also the JT contribution) 
obtained from the exact diagonalization of the first and fourth terms in Eq.(1); (2) 
the SE energy $E_{\rm J}$ directly calculated from the second and third terms in Eq.(1). 
The elastic energy of the JT lattice distortions will not be taken into consideration since the $Q_2$ and $Q_3$ degrees of freedom are fixed in our variational method (therefore the last term of Eq.(1) is just the same constant for all candidate phases). For more details of the Hamiltonian and numerical methods, readers should consult Ref.~\cite{Dagotto:Prp}.

\section{Two-dimensional results}
In our recent publication \cite{Dong:Prb08.2}, the 2D phase diagram was studied, explaining the origin of the realistic spiral order and A-S-E transition in $R$MnO$_3$. The main results can be summarized as follows:

(1) With only the DE plus NN SE interactions (the 1st-2nd terms of Hamiltonian Eq.~1), there is no spiral phase existing between the A-AFM and E-AFM, agreeing with previous studies \cite{Hotta:Prl}.

(2) By considering the DE, NN SE and NNN SE interactions (the 1st-3th terms of Hamiltonian Eq.~1), the SSO phase emerges between the A-AFM and E-AFM when the NNN SE $J_{2b}$ is larger than $0.017t_0$, as shown in Fig. 2(a). Therefore, the A-S-E transition can be qualitatively understood as the enhancement of $J_{\rm AF}/t_0$ and $J_2/t_0$ by the GdFeO$_3$ distortion. However, the required $J_2$ for the SSO remains too large and the obtained $q$ for SSO is lower than the real value in $R$MnO$_3$.

(3) To reproduce the realistic SSO, the whole Hamiltonian (Eq.~1) should be considered, including the contribution from JT distortions. With a modest JT distortion coupling ($\lambda|Q_{2}|=1.5$), the SSO phase region is expanded in parameter space, as shown in Fig. 2(b-c)  \cite{Dong:Prb08.2}. The realistic short-wavelength SSO in $R$MnO$_3$ can be obtained with a weak $J_{2b}$ ($\sim10\%J_{\rm AF}\sim1\%t_0$). In addition, the JT distortion contributes to the insulating nature of $R$MnO$_{3}$, as show in Fig. 2(d), which is crucial for the FE polarization.

(4) The phase transition between the A-AFM and SSO phases is second-order because the wave vector $q$ changes continuously from $0$ (A-AFM) to a finite value (SSO). In contrast, the S-E and A-E phase transitions are of first-order. These orders of the phase transitions are independent of the JT distortions. Interesting physical phenomena, such as the bicritical point and phase separation, may emerge in the vicinity of 
the S-E phase boundary.

(5) The Monte Carlo (MC) simulation on a $12\times12$ lattice confirms the stability of the spiral phase and A-S-E transition at low temperature. The temperature dependent FE polarization is also obtained using the phenomenological equation $-\textbf{e}_{i,j}\times(\textbf{S}_{i}\times\textbf{S}_{j})$, which agrees with experimental observation qualitatively. 
Note here that a finite-size lattice is used in our MC simulations, and those finite lattices 
can only accommodate some particular SSO, such as the state with $q=1/6$ wavevector 
(corresponding to Tb$_{0.41}$Dy$_{0.59}$MnO$_{3}$ \cite{Arima:Prl}). In particular, 
the $12\times12$ lattice is a very good choice since it is compatible with the A-AFM and E-AFM states, 
as well as the $q=1/6$ SSO simultaneously. At present, MC simulation results 
on larger 2D or 3D lattices is  not available due to the rapid growth of  CPU time with increasing lattice sizes. Therefore, in this Proceeding, we will focus most of our attention on the ground states using the zero-$T$ variational method. Readers can consult our original publication (Ref.~\cite{Dong:Prb08.2}) if they are interested in the finite-temperature results on the 2D $12\times12$ lattice.

Note that this weak NNN SE interaction ($<10\%$ NN SE), while shown here to be crucial in the context of the manganite multiferroics, it does not alter the previous large body of investigations and conclusions reached via MC simulations for undoped and doped LaMnO$_3$, since in that case the extra NNN SE couplings can be neglected.

\begin{figure}
\centering \resizebox{0.5\textwidth}{!}{\includegraphics{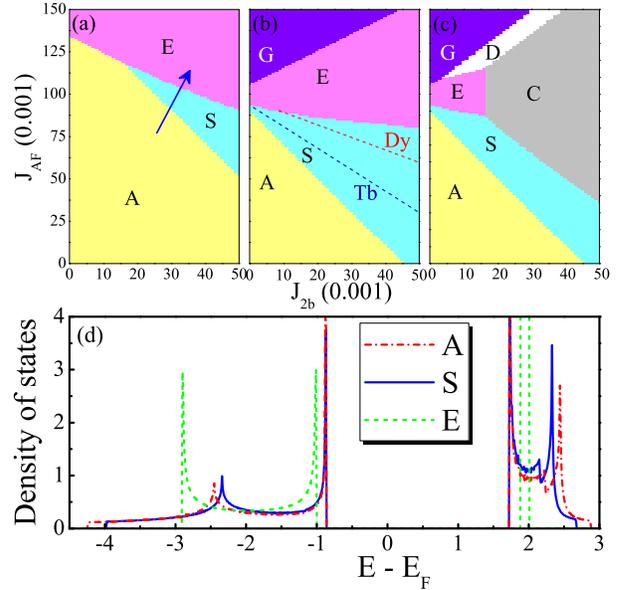}} \caption{(Color online) (a-c) Zero-T phase diagram of the 2D two-orbital DE model for $R$MnO$_{3}$. Notations: A: A-AFM; C: C-AFM; D: Dimer; E: E-AFM; G: G-AFM; S: Spiral spin state. (a) Results without the JT distortion. The possible path for the A-S-E phase transition is indicated by the arrow. The phase diagram is independent of $J_{\rm 2a}$, as long as $J_{\rm 2a}<J_{\rm 2b}$. (b) With the JT phonon ($\lambda |Q_2|=1.5$) and $J_{2a}=0$. The $q$'s for TbMnO$_3$ and DyMnO$_3$ are also indicted by broken lines. (c) The same with (b) except $J_{2a}=0.5J_{2b}$. In (a-c), all the slopes of the A-S boundaries are $-2$ since the SE energy $E_{\rm J}$ 
for the A-AFM and SSO phases is $2J_{\rm AF}\cos(\delta)+J_{\rm 2b}\cos(2\delta)+J_{\rm 2a}$, where $\delta$ is the NN spin angle. (d) Density of states for the A-AFM, SSO ($q=0.14$ as TbMnO$_3$), and E-AFM with the same JT distortion. (a-c) Reproduced from \cite{Dong:Prb08.2}; Copyright 2008, the American Physical Society.}
\end{figure}

\section{Three-dimensional results}
In the above 2D study, the ferromagnetic (FM) phase was neglected in the candidate list. This was reasonable since none of the $R$MnO$_3$ bulk materials is FM. However, very recently, the FM tendency in LaMnO$_3$ and TbMnO$_3$ thin films grown on SrTiO$_3$ substrate has been observed \cite{Bhattacharya:Prl,Rubi:Prb,Kirby:Jap}. To fully understand this exotic behavior, a 3D lattice is necessary. In addition to the candidate phases in the above 2D studies, two more phases: the FM and 3D canting spin phase are included. The 3D canting spin order is similar to the forementioned SSO, namely both of them are noncollinear spin orders and can be scaled by wave vectors $q$. However, the SSO is only noncollinear in the $a-b$ plane, while spins are collinear along $c$ axis. In contrast, in the 3D canting spin order, spins are noncolinear along all the directions, namely the NN spins' angles are isotropic. To simplify the model, all the $12$ NNN SE couplings are firstly assumed to be equal.

Using the same zero-T variational method, we first calculate the phase diagram on an infinite cubic lattice without the JT distortion, as shown in Fig. 3(a). Comparing with Fig.~2(a), there are several differences. In the 3D lattice phase diagram, the FM phase occupies a large region at low $J_{\rm AF}$ and $J_2$. The A-AFM can not exist until $J_2>0.018$. This FM-rich phase diagram can shed light to the FM tendency in $R$MnO$_3$ thin films. The LaMnO$_3$ lattice on a SrTiO$_3$ substrate is almost cubic \cite{Bhattacharya:Prl,Dong:Prb08.3}. Even for the TbMnO$_3$ thin film, the differences between $a$, $b$, and $c/\sqrt{2}$ are also reduced by the stress \cite{Rubi:Prb,Kirby:Jap}, namely the thin-film lattice becomes closer to cubic than that of bulk. This FM enhancement in the cubic lattice can be understood based on three observations: 
(1) the isotropic exchange interactions due to the isotropic Mn-O-Mn bond-length and bond-angle do not prefer phases with anisotropic spin patterns, e.g. A-AFM; (2) the Mn-O-Mn bond-angle is more straight in the higher-symmetry lattice, which will enhance the ratio between the DE and SE terms, giving rise to an increased bandwidth and FM tendency; (3) the orthorhombic distortion in the bulk is advantageous for the $d_{\rm 3x^2-r^2}$/$d_{\rm 3y^2-r^2}$ type orbital-ordering in the A-AFM and SSO phases, but this orbital order 
will be suppressed in the nearly cubic lattice on the substrate. In summary, the FM tendency in $R$MnO$_3$ is natural once the lattice is close to cubic.

Similar to the 2D case, to understand the properties of $R$MnO$_3$ in the bulk, the JT distortion of orthorhombic lattice has to be considered. Using the same $\lambda |Q_2|=1.5$, the phase diagram is recalculated, as shown in Fig.~3(b). The FM region is largely suppressed, and it is being replaced by a robust A-AFM phase, which agrees with the properties of  real $R$MnO$_3$ ($R=$La, Pr, Nd, ...) bulk materials. The $q$'s for TbMnO$_3$ is also indicted. However, the spiral phase region is somehow narrow, and the realistic $q$'s for DyMnO$_3$ are missing. To solve this puzzle, the anisotropy of NNN SE has to be considered \cite{Dong:Prb08.2}. Since the accurate ratios between $J_2$'s along different directions are unclear, here only two limits are calculated. In Fig.~3(a-b), the upper limit, namely isotropic $J_2$'s, has been considered. 
With the isotropic $J_2$, six equal NNN SE bonds per site are taken into account. For 
the lower limit, only the $J_{\rm 2b}$ is nonzero (it is along the $b$-direction because it has 
the strongest intensity due to the largest Mn-O-O-Mn angle), while other $J_2$s are all set to be zero. 
In this limit case, only one NNN SE bond per site is considered. The new phase diagram is calculated and shown in Fig.~3(c). The spiral phase region is expanded in this revised phase diagram, and now the wave vector $q$ for DyMnO$_3$ can be found in Fig.~3(c). In real manganites, the $J_2$ should be within 
these two limits, namely six inequivalence NNN SE bonds per site should be considered.

In short, the 3D calculation agrees with the 2D results qualitatively. Besides the SSO and the A-S-E transition, the FM tendency in $R$MnO$_3$ thin films is also explained by considering the substrate stress. The several phases existing in the phase diagram illustrate the possibility to modulate the subtle phase competition in $R$MnO$_3$ using various methods, such as stress and strain, or external magnetic and electric fields. With these stimulations, phase separation may be possible to emerge, which would result in colossal responses to these external stimulations.

\begin{figure}
\centering
\resizebox{0.5\textwidth}{!}{\includegraphics{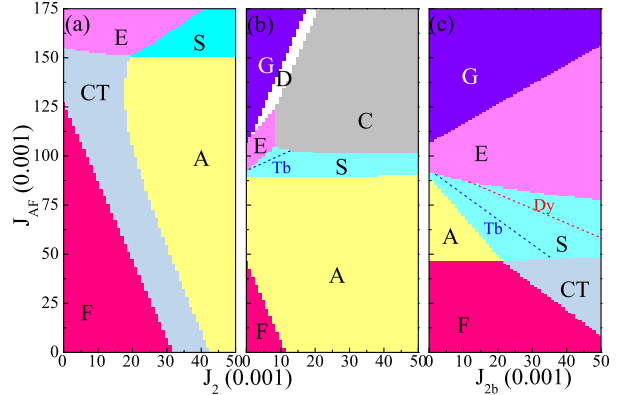}}
\caption{(Color online) Zero-T phase diagram of the 3D two-orbital DE model for $R$MnO$_{3}$. Notations: F: FM; CT: 3D canting spin order; others conventions are the same as those in Fig.~2(a-c). (a) Results without the JT distortion and with the isotropic $J_2$ coupling. (b) Results with the JT distortion ($\lambda |Q_2|=1.5$) and with the isotropic $J_2$ coupling. The $q$'s for TbMnO$_3$ are indicated. (c) Same as (b) except that the anisotropic $J_2$ is used. Only the $J_{2b}$ is nonzero while others are zero.}
\end{figure}

\section{More Discussion}
In our publication \cite{Dong:Prb08.2}, we argued that although the classical spin model with NNN magnetic frustration provides a natural starting point to describe phenomenologically the spiral phase, this simple model is not sufficient to deeply understand the microscopic origin of the SSO in perovskite manganites. In the $J_{1}$-$J_{2}$-$J_{3}$ model with classical spins, a strong $J_{2}$ (compared with $J_{1}$) is required to generate a realistic short-wavelength SSO, e.g. to reach the realistic wave vectors of TbMnO$_3$ and DyMnO$_3$, the $J_{2}/|J_{1}|$ should be about $0.78\sim1.35$ \cite{Kimura:Prb}. However, the first-principles calculations show that the value of $J_{2}$  is only about $0.56|J_{1}|$ \cite{Xiang:Prl}, which is not sufficient to induce the experimentally observed short-length ($L=6$) SSO. Also, the coupling $J_{3}$ was found to be $0.375|J_{1}|$ and in addition it is AFM, opposite to the weak FM coupling ($0.01J_{1}$) used in the previous model study \cite{Kimura:Prb}. Furthermore, the classical spin model is not suitable to explore in a single framework the several phases (A-AFM, spiral, and E-AFM) found in $R$MnO$_3$, e.g. it can not distinguish the E-AFM order from the $q=1/4$ spiral order since they have the same energy.

Very recently, the electromagnon excitations in $R$MnO$_3$ were studied \cite{Katsura:Prl07,Aguilar:cm}. The dynamics of the spiral order is a very important issue, which can be helpful to discover the real mechanism of multiferroicity. The $J_{1}$-$J_{2}$-$J_{3}$ model with classical spins was used to calculate the spin wave properties \cite{Aguilar:cm}. Thus, it is necessary to clarify the properties of the classical spin model further to decide whether such an approach is sufficient to study the spin waves in $R$MnO$_3$.

Since the $e_g$ density is uniform for $R$MnO$_3$, in the following we can use a classical approximation for the DE interaction \cite{Tsai:Jap}:
\begin{equation}
H=\sum_{<ij>}[J_{\rm DE}\sqrt{1+\textbf{S}_i\cdot\textbf{S}_j}+J_{\rm AF}\textbf{S}_i\cdot\textbf{S}_j]+\sum_{[ij]}J_{2\gamma}\textbf{S}_{i}\cdot\textbf{S}_{j},
\end{equation}
where $J_{\rm DE}$ is an effective coupling which can be derived from the kinetic energy. \emph{Note that this approximation gives the same energy as in the full study of the DE model only when the NN spins' couplings are uniform. Also, this approximation is accurate only when the JT distortions are neglected.} For instance, $J_{\rm DE}$ is about $-0.423t_{0}$ for the 2D A-AFM or the spiral order in $R$MnO$_3$ since the DE kinetic energy (per site) is about $-1.196t_{0}$. But for the 3D FM and canted states of $R$MnO$_3$, $J_{\rm DE}$ is about $-0.363t_{0}$ since the DE kinetic energy (per site) is about $-1.541t_{0}$. In the following, we will only focus our attention on the 2D SSO phase.

Using the standard expansion technique and rotation of the quantum projection axis, the linear spin-wave theory can be applied to the SSO phase \cite{Merino:Jpcm}. It is straightforward to obtain an equivalent NN coupling $J_1=\frac{J_{\rm DE}}{2\sqrt{1+\cos(\delta)}}+J_{\rm AF}$, where $\delta$ is the (ground state) angle between NN the spins. This coupling $J_1$ is weaker than $t_0$ and $J_{\rm AF}$, and comparable with $J_2$. Therefore, the classical $J_1$-$J_2$-$J_3$ model appears suitable for this simple description to some extent.

Even when the JT distortion is included, although the exact formula is unknown the classical DE term can still merge with the NN SE term, resulting in a weak effective $J_1$ coupling. Therefore, it seems possible to use the $J_{1}$-$J_{2}$-$J_{3}$ model with classical spins to study the spin wave in the SSO $R$MnO$_3$. However, once there are phases with nonuniform NN spins couplings (e.g. E-AFM and CE phases), 
the DE Hamiltonian can not be simplified into a mere classical Heisenberg model, thus several phases in manganites can not be accessed by using a pure spin model. The physical reason is that the $e_{\rm g}$ electrons are itinerant (due to the DE hopping) even in the undoped $R$MnO$_3$, while the pure spin model is based entirely on localized spins. This itinerant DE process is essential to understand the novel physics in manganites, for instance the charge/orbital ordering in the zigzag chains of the E-AFM or CE phases \cite{Dagotto:Prp,Hotta:Rpp}, and also the multiferroicity in the E-AFM state of 
HoMnO$_3$ \cite{Sergienko:Prl}. Thus, it is not correct to investigate the existence of several competing phases in $R$MnO$_3$, such as the phase diagram and phase transitions, using just a pure spin classical approximation.

\section{Conclusion}
In conclusion, here we have provided a microscopic description of the several competing spin orders in multiferroic $R$MnO$_{3}$ perovskites. The experimentally observed spiral order and FE transition can be obtained by incorporating a weak NNN superexchange interaction and a Jahn-Teller distortion into the standard two-orbitals DE model for manganites. Several aspects of the experimentally known A-S-E phase transition with decreasing $R$ size are well reproduced by including the GdFeO$_{3}$-type distortion in our study. Furthermore, the FM tendency in $R$MnO$_{3}$ thin films is also explained by considering the substrate stress. The relationship between a previously studied classical spin model for multiferroics and our more fundamental model is also further discussed.

\acknowledgement{This work was supported by the NSF (DMR-0706020) and the Division of Materials Science and Engineering, U.S. DOE, under contract with UT-Battelle, LLC. J.M.L. was supported by the National Key Projects for Basic Research of China (2006CB921802 and 2009CB929501) and the National Natural Science Foundation of China (50832002). S. Y. was supported by CREST-JST. S.D. was also supported by the China Scholarship Council.}


\begin{thebibliography}{37}

\bibitem{Fiebig:Jpd}
M.~Fiebig, J. Phys. D: Appl. Phys \textbf{38}, R123 (2005)

\bibitem{Eerenstein:Nat}
W.~Eerenstein, N.D. Mathur, J.F. Scott, Nature (London) \textbf{442}, 759 (2006)

\bibitem{Cheong:Nm}
S.-W. Cheong, M.~Mostovoy, Nature Mater. \textbf{6}, 13 (2007)

\bibitem{Ramesh:Nm}
R.~Rameshi, N.A. Spaldin, Nature Mater. \textbf{6}, 21 (2007)

\bibitem{Kimura:Nat}
T.~Kimura, T.~Goto, H.~Shintani, K.~Ishizaka, T.~Arima, Y.~Tokura, Nature (London) \textbf{426}, 55 (2003)

\bibitem{Goto:Prl}
T.~Goto, T.~Kimura, G.~Lawes, A.P. Ramirez, Y.~Tokura, Phys. Rev. Lett. \textbf{92}, 257201 (2004)

\bibitem{Kenzelmann:Prl}
M.~Kenzelmann, A.B. Harris, S.~Jonas, C.~Broholm, S.B.K. J.~Schefer, C.L. Zhang, S.-W. Cheong, O.P. Vajk, J.W. Lynn, Phys. Rev. Lett. \textbf{95},
  087206 (2005)

\bibitem{Kimura:Prb05}
T.~Kimura, G.~Lawes, T.~Goto, Y.~Tokura, A.P. Ramirez, Phys. Rev. B \textbf{71}, 224425 (2005)

\bibitem{Arima:Prl}
T.~Arima, A.~Tokunaga, T.~Goto, H.~Kimura, Y.~Noda, Y.~Tokura, Phys. Rev. Lett. \textbf{96}, 097202 (2006)

\bibitem{Lorenz:Prb}
L.~Lorenz, Y.Q. Wang, C.W. Chu, Phys. Rev. B \textbf{76}, 104405 (2007)

\bibitem{Hur:Nat}
N.~Hur, S.~Park, P.A. Sharma, J.S. Ahn, S.~Guha, S.-W. Cheong, Nature (London) \textbf{429}, 392 (2004)

\bibitem{Chapon:Prl}
L.C. Chapon, G.R. Blake, M.J. Gutmann, S.~Park, N.~Hur, P.~Radaelli, S.-W. Cheong, Phys. Rev. Lett. \textbf{93}, 177402 (2004)

\bibitem{Cruz:Prb}
C.R. dela. Cruz, B.~Lorenz, Y.Y. Sun, Y.~Wang, S.~Park, S.-W. Cheong, M.M. Gospodinov, C.W. Chu, Phys. Rev. B \textbf{76}, 174106 (2007)

\bibitem{Zhou:Prl}
J.S. Zhou, J.B. Goodenough, Phys. Rev. Lett. \textbf{96}, 247202 (2006)

\bibitem{Mostovoy:Prl}
M.~Mostovoy, Phys. Rev. Lett. \textbf{94}, 137205 (2005)

\bibitem{Katsura:Prl}
H.~Katsura, N.~Nagaosa, A.V. Balatsky, Phys. Rev. Lett. \textbf{95}, 057205 (2005)

\bibitem{Sergienko:Prb}
I.A. Sergienko, E.~Dagotto, Phys. Rev. B \textbf{73}, 094434 (2006)

\bibitem{Xiang:Prl}
H.J. Xiang, S.H. Wei, M.H. Whangbo, J.L.F.D. Silva, Phys. Rev. Lett. \textbf{101}, 037209 (2008)

\bibitem{Malashevich:Prl}
A.~Malashevich, D.~Vanderbilt, Phys. Rev. Lett. \textbf{101}, 037210 (2008)

\bibitem{Kimura:Armr}
T.~Kimura, Annu. Rev. Mater. Res. \textbf{37}, 387 (2007)

\bibitem{Sergienko:Prl}
I.A. Sergienko, C.~\c{S}en, E.~Dagotto, Phys. Rev. Lett. \textbf{97}, 227204 (2006)

\bibitem{Picozzi:Prl}
S.~Picozzi, K.~Yamauchi, B.~Sanyal, I.A. Sergienko, E.~Dagotto, Phys. Rev. Lett. \textbf{99}, 227201 (2007)

\bibitem{Yamauchi:Prb}
K.~Yamauchi, F.~Freimuth, S.~Bl\"{u}gel, S.~Picozzi, Phys. Rev. B \textbf{78}, 014403 (2008)

\bibitem{Hotta:Prl}
T.~Hotta, M.~Moraghebi, A.~Feiguin, A.~Moreo, S.~Yunoki, E.~Dagotto, Phys. Rev. Lett. \textbf{90}, 247203 (2003)

\bibitem{Kimura:Prb}
T.~Kimura, S.~Ishihara, H.~Shintani, T.~Arima, K.T. Takahashi, K.~Ishizaka, Y.~Tokura, Phys. Rev. B \textbf{68}, 060403(R) (2003)

\bibitem{Li:Prb}
Q.~Li, S.~Dong, J.M. Liu, Phys. Rev. B \textbf{77}, 054442 (2008)

\bibitem{Dong:Prb08.2}
S.~Dong, R.~Yu, S.~Yunoki, J.M. Liu, E.~Dagotto, Phys. Rev. B \textbf{78}, 155121 (2008)

\bibitem{Dagotto:Prp}
E.~Dagotto, T.~Hotta, A.~Moreo, Phys. Rep. \textbf{344}, 1 (2001)

\bibitem{Zhou:Prb}
J.S. Zhou, J.B. Goodenough, Phys. Rev. B \textbf{77}, 132104 (2008)

\bibitem{note}
In $R$MnO$_3$, the order of the Mn$^{3+}$ spins can be characterized by a particular propagation vector ($0$, $q_{\rm Mn}$, $1$) (in the orthorhombic $Pbnm$ cell notation, see Fig.~1(a), indicating an AFM coupling along the $c$ axis, and FM coupling along the $a$ axis. $q_{\rm Mn}$ is $0$ for the A phase and $0.5$ for the E phase. However, for convenience from the theoretical viewpoint, $q$ in the following will be defined along the $x/y$ directions (see Fig.~1(a)) which equals half of the $q_{\rm Mn}$ usually used in the experimental papers.

\bibitem{Bhattacharya:Prl}
A.~Bhattacharya, S.J. May, S.G. te~Velthuis, M.~Warusawithana, X.~Zhai,  B.~Jiang, J.M. Zuo, M.R. Fitzsimmons, S.D. Bader, J.N. Eckstein, Phys. Rev.  Lett. \textbf{100}, 257203 (2008)

\bibitem{Rubi:Prb}
D.~Rubi, C.~de~Graaf, C.J.M. Daumont, D.~Mannix, R.~Broer, B.~Noheda, Phys.  Rev. B \textbf{79}, 014416 (2009)

\bibitem{Kirby:Jap}
B.J. Kirby, D.~Kan, A.~Luykx, M.~Murakami, D.~Kundaliya, I.~Takeuchi, J. Appl. Phys. \textbf{105}, 07D917 (2009)

\bibitem{Dong:Prb08.3}
S.~Dong, R.~Yu, S.~Yunoki, G.~Alvarez, J.M. Liu, E.~Dagotto, Phys. Rev. B  \textbf{78}, 201102(R) (2008)

\bibitem{Katsura:Prl07}
H.~Katsura, A.V. Balatsky, N.~Nagaosa, Phys. Rev. Lett. \textbf{98}, 027203  (2007)

\bibitem{Aguilar:cm}
R.V. Aguilar, M.~Mostovoy, A.B. Sushkov, C.L. Zhang, Y.J. Choi, S.-W. Cheong,  H.D. Drew, Phys. Rev. Lett. \textbf{102}, 047203 (2009)

\bibitem{Tsai:Jap}
S.H. Tsai, D.P. Landau, J. Appl. Phys. \textbf{87}, 5807 (2000)

\bibitem{Merino:Jpcm}
J.~Merino, R.H.M. Kenzie, J.B. Marston, C.H. Chung, J. Phys.: Condens. Matter  \textbf{11}, 2965 (1999)

\bibitem{Hotta:Rpp}
T.~Hotta, Rep. Prog. Phys. \textbf{69}  2061 (2006)

\end{thebibliography}

\end{document}